\documentclass[aps,graphicx,twocolumn,prd,eqsecnum,showpacs]{revtex4-1}
\usepackage{graphicx}
\usepackage{dcolumn}
\usepackage{bm}
\begin{document}
\title{Bose gas in power-like spherically symmetric potential in
arbitrary spatial dimensionality.}

\author{A.~A.~Kozhevnikov}
\email[]{kozhev@math.nsc.ru} \affiliation{Laboratory of
Theoretical Physics, S.~L.~Sobolev Institute for Mathematics, and
Novosibirsk State University, 630090, Novosibirsk, Russian
Federation}

\date{\today}
\begin{abstract}
Temperature of the  Bose -- Einstein condensation and the
temperature behavior of the chemical potential and other
thermodynamical functions of the ideal Bose gas are found for the
arbitrary power-like spherical-symmetric potential at an arbitrary
space dimension. It is shown that the recently observed Bose --
Einstein condensation of photons in the cavity is the phase
transition of the third kind.
\end{abstract}

\maketitle

\section{Introduction}
~

The  majority  of studies, both theoretical and experimental, on
the Bose -- Einstein condensation (BEC) was carried out for gases
in the harmonic traps and in three space dimensions \cite{Pit}. In
the meantime, a number of works were devoted to the study of the
thermodynamic properties of the Bose gas trapped  in arbitrary
power-like spherical-symmetric potential in three spacial (3D)
dimensions \cite{Bag}, and to the harmonic trap at lower dimension
$D=2$ \cite{Mull}.  Recent publication \cite{achar} contains the
calculation  of the temperature and the fraction of the condensed
particles in case of Bose -- Einstein condensation of the ideal
Bose gas in free space at any space dimension. As far as the
physical realization is concerned, the harmonic traps of lower
dimensions are effectively modeled by the sharply anisotropic
frequencies \cite{Mull}. A direct realization of BEC in the 2D
harmonic and  isotropic trap  was recently achieved with the
photons confined in the spherical cavity \cite{Kla}. The
confinement results both in the effective nonrelativistic
dependence of the photon energy on the transverse momentum, the
harmonic oscillator potential in the transverse direction,
\begin{eqnarray}
E&=&\hbar c\sqrt{k^2_\parallel(r)+k_r^2}\approx\hbar
ck_\parallel(r) +\frac{\hbar
ck^2_r}{2k_\parallel(r)}\approx\nonumber\\&& \approx\frac{\pi\hbar
cs}{D_0}+\frac{\pi\hbar csr^2}{RD^2_0}+\frac{\hbar cD_0k^2_r}{2\pi
s}\equiv\nonumber\\&&E_0+\frac{m_{\rm
eff}\omega^2r^2}{2}+\frac{p^2_r}{2m_{\rm eff}},
\label{Edisp}\end{eqnarray} with a tiny effective photon mass
$m_{\rm eff}=\pi\hbar cs/D_0\approx5\times10^{-33}$ gramm, the
frequency $\omega=c\sqrt{2/(RD_0)}\approx4\times10^{11}$ s$^{-1}$,
and $E_0=\pi\hbar cs/D_0\approx3$ eV \cite{Kla}. Here,
$D_0=1.4\times10^{-4}$ cm is the distance along the optical axes
between the spherical mirrors, each having the radius of curvature
$R=100$ cm, $s=7$ is the excitation number of the longitudinal
standing wave of the photon, and $r$, $k_r=p_r/\hbar$ are,
respectively, the distance from the optical axes and the conjugate
wave number in the transverse direction, $c$ and $\hbar$ are the
velocity of light and the Planck constant.

However, the general analysis of the possibility of BEC in a
power-law potential in an arbitrary space dimension $D$, as well
as the degree of singularity of the chemical potential and other
thermodynamic functions, is absent in the literature. In the
present work we fill the gap and find  both the conditions for
BEC, as well as the degree of singularity of the thermodynamical
functions of the ideal Bose gas, appearing in the case of
arbitrary power-like spherical potential and for arbitrary space
dimensionality. Despite the fact that the ideal Bose gas
approximation adopted in the present work is poor for the gases of
alkali metals, it is meaningful for  BEC of photons studied in
Ref.~\cite{Kla}, in addition to the case of the  hypothetical
weakly interacting axions, proposed as  dark-matter candidate in
Ref.~\cite{Sik}.

The rest of the paper is organized as follows. The density of
states is obtained in  Section \ref{densst}. Sec.~\ref{thermf} is
devoted to the elucidation of the behavior of the chemical
potential near the the BEC transition temperature. The results and
conclusions are presented in Sec.~\ref{disc}. Appendix contains
the necessary information concerning the polylogarithm function.

\section{Density of states and the critical temperature of BEC}
\label{densst}
~

The starting point is the semi-classical  one-particle density of
states $\nu(\epsilon)$ in the case of  spherically symmetric
single-particle potential of the form
\begin{equation}
U(r)=U_0\left(\frac{r}{a}\right)^\gamma,\label{Ur}\end{equation}
characterized by its strength $U_0>0$, the scale $a$, and the
exponent $\gamma$. In effect, the single-particle density of
states can be defined as
\begin{equation}
\nu(\epsilon)=\int\frac{d^Drd^Dp}{(2\pi\hbar)^D}\delta\left[\epsilon-\frac{p^2}{2m}-
U_0\left(\frac{r}{a}\right)^\gamma\right],\label{nu}\end{equation}where
$\delta(x)$ is the Dirac delta function. Hereafter $D$ is the
space dimensionality of the gas, the particle spin is assumed to
be zero, and $p$ and $r$ are the moduli of the $D$-dimensional
vectors of the momentum and the radius, respectively. The
phase-space integral (\ref{nu}) is calculated in the
multi-dimensional spherical coordinates, where the volume element
integrated over angular variables takes the form
$$d^Dr=\frac{2\pi^{D/2}}{\Gamma\left(\frac{D}{2}\right)}r^{D-1}dr.$$
See \cite{ryzh}, the formula 4.632(2) differentiated over $R$.
Analogously for  $d^Dp$. With the use of the table integral
\cite{grad1}
$$\int_0^1x^{\xi-1}(1-x)^{\eta-1}dx=\frac{\Gamma(\xi)\Gamma(\eta)}{\Gamma(\xi+\eta)},$$
the calculation leads to the explicit expression
\begin{equation}
\nu(\epsilon)=A\epsilon^{\alpha-1},\label{nuBEC}\end{equation}
where
\begin{eqnarray}
\alpha&=&\frac{D}{2}+\frac{D}{\gamma},\nonumber\\
A&=&\left(\frac{ma^2}{2\hbar^2U_0^{2/\gamma}}\right)^{D/2}
\frac{2\Gamma\left(\frac{D}{\gamma}\right)}{\gamma\Gamma\left(\frac{D}{2}\right)
\Gamma\left(\alpha\right)},\label{notat}\end{eqnarray} and
$\Gamma(x)$ is the Euler gamma function.

The case of $\gamma=0$, at first sight, seems to be singular.
However, it corresponds to the situation, when the gas is  under
the influence of the constant potential $U_0$. As is usual in the
case of free particles or particles in the constant potential, one
should  put the system  into the finite volume. The problem is
reduced to the case of bosons in $D$ spatial dimensions, with the
only difference that they  condense to the state with the energy
$U_0$. The temperature of BEC in this case is the same as found in
\cite{achar}. The thermodynamical properties of the system do not
depend on changing the energy reference point.

In the meantime, there is another way to get the case of free
bosons in the space with no external potential,  confined inside
the sphere with the radius $a$. This is obtained in the limit
$\gamma\to\infty$. Then
\begin{equation}
U(r)=U_0\left(\frac{r}{a}\right)^\gamma\to\left\{\begin{array}{c}
           0\mbox{, if }r<a, \\
           \infty\mbox{, if } r>a,\\
         \end{array}\right.\label{gamminf}\end{equation}
and the expressions (\ref{notat}) are reduced, respectively, to
the following ones
\begin{eqnarray}
\alpha&=&\frac{D}{2},\nonumber\\
A&=&\left(\frac{m}{2\hbar^2}\right)^{D/2}
\frac{a^D}{\Gamma\left(\frac{D}{2}\right)
\Gamma\left(\frac{D}{2}+1\right)}.\label{notat1}\end{eqnarray} The
density of states in this limit coincides with that  obtained
earlier \cite{achar} for free bosons in a finite volume. As for
the dependence of the chemical potential on the temperature, in
the cases of $\gamma=0$ or $\gamma\to\infty$, we will comment on
this point after establishing the analogous dependence in the
general case of finite  $\gamma\not=0$.

As usual, the inequalities $\mu(T)\leq0$ and
$\left(\frac{\partial\mu}{\partial T}\right)_{U_0}<0$, result from
the condition of having the  particle number bosons in the gas,
i.e.,
\begin{equation}
N=\int_0^\infty\frac{\nu(\epsilon)d\epsilon}{e^{(\epsilon-\mu)/T}-1},
\label{muviaN}\end{equation}where $\mu\equiv\mu(T)$ is the
chemical potential. The above equation is valid only if $U_0>0$.
The Boltzmann constant is set to unity. BEC is possible if the
integral defining the temperature of BEC $T_0$,
\begin{eqnarray}
N&=&\int_0^\infty\frac{\nu(\epsilon)d\epsilon}{e^{\epsilon/T_0}-1}=
\left(\frac{ma^2}{2\hbar^2U_0^{2/\gamma}}\right)^{D/2}\times\nonumber\\&&
\frac{2\Gamma\left(\frac{D}{\gamma}\right)\zeta\left(\alpha\right)T_0^{\alpha}}
{\gamma\Gamma\left(\frac{D}{2}\right)},\label{NTBEC}\end{eqnarray}has
a definite value. Hereafter
$$\zeta(x)=\sum_{n=1}^\infty\frac{1}{n^x}$$ is the Riemann zeta
function. This function is defined at $x>1$. Hence, the
restriction on the parameters of the problem $\gamma$ and $D$,
resulting from the definition of zeta function, is
\begin{equation}
\alpha=\frac{D}{2}+\frac{D}{\gamma}>1.\label{cond1}\end{equation}
One obtains that BEC is possible in the space dimension  $D=1,2$,
and 3 at, respectively, $\gamma<2$, $\gamma>0$, and $\gamma>-6$.
In particular, the confinement in the  harmonic trap,  $\gamma=2$,
admits BEC in any space dimension when  $D>1$. The critical
temperature $T_0$ of BEC is found from Eq.~(\ref{NTBEC}):
\begin{equation}
T_0=\left[N\left(\frac{2\hbar^2U_0^{2/\gamma}}{ma^2}\right)^{D/2}\frac{\gamma
\Gamma\left(\frac{D}{2}\right)}{2\Gamma\left(\frac{D}{\gamma}\right)\zeta(\alpha)}\right]^{1/\alpha}.
\label{T0}\end{equation} In particular, one has $D=3$, $\gamma=2$
for BEC of alkali atoms in the traps with the typical magnitude of
the frequency $\omega\sim10^2$ s$^{-1}$. Then
$ma^2/(2\hbar^2U_0)=(\hbar\omega)^{-2}$, and the BEC temperature
is
$$T_0=\hbar\omega\left[\frac{N}{\zeta(3)}\right]^{1/3}\sim10^{-7}\mbox{
K}$$ for $N\sim 10^6$. Hereafter, the necessary expressions for
the zeta function at particular arguments  are taken from
\cite{LL5}. On the other hand, under the conditions of the
experiment \cite{Kla}, the parameters are $D=2$, $\gamma=2$,
$\omega\sim10^{11}$ s$^{-1}$, and  the BEC temperature is
$$T_0=\frac{\hbar\omega}{\pi}\sqrt{3N}\sim10^3\mbox{ K},$$ at
$N\sim10^4$ photons in the cavity. The above equation allows for
two polarization states of the photon. For the oscillator
potential in space with arbitrary  dimension $D$ one finds
\begin{equation}
T_0=\hbar\omega\left[\frac{N}{\zeta(D)}\right]^{1/D}.\label{T0D}\end{equation}
Since lim$_{D\to\infty}\zeta(D)=1$, one can see from (\ref{T0D})
that, in this limit, $T_0\to\hbar\omega$ independently of the
particle number $N$.

The chemical potential $\mu(T)$ at  $T<T_0$ is zero with all its
derivatives. Hence, the number of particles sitting at the excited
energy levels, $N_>(T)$, is given by Eq.~(\ref{NTBEC}), in which
one should make the replacements $N\to N_>(T)$ and  $T_0\to T$.
Then, the number of particles in BEC, $N_0(T)=N-N_>(T)$, is given
by the expression
\begin{equation}
N_0(T)=N\left[1-\left(\frac{T}{T_0}\right)^\alpha\right].\label{N_0}\end{equation}
This expression is valid for the finite values of the power
$\gamma$ in the external potential. It coincides with the earlier
expression $\alpha=3/2+3/\gamma$ \cite{Pit,Bag} at $D=3$. As is
explained above, the case of ideal Bose gas put in a finite
volume, upon neglecting the  interaction with the external field,
corresponds to taking the limit $\gamma\to\infty$, and the power
$\alpha$ is reduced to the value $\alpha=3/2$ in the
three-dimensional case.

\section{Thermodynamic functions near BEC transition temperature}
\label{thermf}
~

The temperature behavior of the thermodynamic functions in the
vicinity of the critical temperature $T_0$ is found from the
behavior of $\mu(T)$. Since $\mu(T)=0$ at $T\leq T_0$, one should
consider the region $T=T_0+\Delta T$, where $\Delta T\ll T_0$.
$\mu(T)$ is found from Eq.~(\ref{muviaN}):
\begin{equation}
\left(\frac{T_0}{T}\right)^\alpha=\frac{{\rm
Li}_\alpha\left(e^{\mu/T}\right)}{{\rm
Li}_\alpha(1)}\label{muexact},\end{equation}where Li$_\alpha(z)$
is  the polylogarithm function. Its definition and some properties
necessary in the context of the present work, are gathered in
Appendix. With the help of Eq.~(\ref{Li}) in the Appendix one can
find the internal energy  in terms of $\mu\equiv\mu(T)$. To this
end one should use the following relations:
\begin{eqnarray*}
E&=&A\int_0^\infty\frac{\epsilon^\alpha
d\epsilon}{e^{(\epsilon-\mu)/T}-1}=AT^{\alpha+1}\Gamma(\alpha+1){\rm
Li}_{\alpha+1}\left(e^{\mu/T}\right),\nonumber\\
N&=&A\int_0^\infty\frac{\epsilon^{\alpha-1}
d\epsilon}{e^{(\epsilon-\mu)/T}-1}=AT^\alpha\Gamma(\alpha){\rm
Li}_{\alpha}\left(e^{\mu/T}\right).\label{EN}\end{eqnarray*}
Excluding from these equations the multiplicative factor $A$, one
obtains the internal energy in the following form:
\begin{equation}
E=NT\alpha\frac{{\rm Li}_{\alpha+1}\left(e^{\mu/T}\right)}{{\rm
Li}_\alpha\left(e^{\mu/T}\right)}. \label{E}\end{equation}
Although Eqs.~(\ref{muexact}) and (\ref{E})  are exact, they are
not very useful for practical purposes. One should consider them
in various limiting cases, in order to obtain  explicit
expressions for $\mu$ and other thermodynamic functions. As will
be clear, the temperature behavior of $\mu(T)$ is different at
different values of $\alpha$. If $\alpha>2$ (that is, if
$\gamma<2/3$, $\gamma<1$, and $\gamma<6$ at $D=1$, $D=2$, and
$D=3$, respectively)  then one may use power expansion in fugacity
$e^{\mu/T}$:
\begin{eqnarray}
N&=&A\int_0^\infty\frac{\epsilon^{\alpha-1}d\epsilon}{e^{(\epsilon-\mu)/T}-1}=
A\sum_{n=0}^\infty e^{\mu(n+1)/T}\int_0^\infty
d\epsilon\times\nonumber\\&&\epsilon^{\alpha-1}
e^{-\epsilon(n+1)/T}=A\Gamma\left(\alpha\right)T^{\alpha}\times\nonumber\\&&
\left[\zeta\left(\alpha\right)+\sum_{n=1}^\infty\frac{e^{\mu
n/T}-1}{n^{\alpha}}\right]. \label{muviaN1}\end{eqnarray} Since
$\mu\ll T$, which is valid in the vicinity of  $T_0$, then, to the
first order, $e^{\mu n/T}-1\approx\mu n/T$, and the summation is
obtained  in the following closed form:
$$\frac{\mu}{T}\sum_{n=1}^\infty\frac{1}{n^{\alpha-1}}=\frac{\mu}{T}
\zeta\left(\alpha-1\right).$$ The explicit equation for finding
the chemical potential valid at $\alpha>2$ is
$$N=N\left(\frac{T}{T_0}\right)^{\alpha}\left[1+\frac{\mu\zeta\left(\alpha-1\right)}
{T\zeta\left(\alpha\right)}\right],$$ whose solution at $T>T_0$ is
\begin{eqnarray*}
\mu(T)&\approx&T_0\left[\left(\frac{T_0}{T}\right)^{\alpha}-1\right]
\frac{\zeta\left(\alpha\right)}{\zeta\left(\alpha-1\right)}\approx
-(T-T_0)
\frac{\alpha\zeta\left(\alpha\right)}{\zeta\left(\alpha-1\right)}.
\end{eqnarray*}One can see from this expression that $\mu(T)$ is continuous at
$T_0$, but its first derivative has a discontinuity
$$[\mu^\prime]=-
\frac{\alpha\zeta\left(\alpha\right)}{\zeta\left(\alpha-1\right)}.$$Hereafter
$$[f]\equiv{\rm lim}_{\epsilon\to0}\left[f(T_0+\epsilon)-f(T_0-\epsilon)
\right]$$ designates the discontinuity of the function $f(T)$ at
the BEC transition temperature $T_0$. Hence, the energy of the
Bose gas is continuous, but the heat capacity at constant external
field defined as
\begin{widetext}
$$C=\left(\frac{\partial E}{\partial T}\right)_{N,U}=A\int_0^\infty\frac{\epsilon^{\alpha}e^{(\epsilon-\mu)/T}}
{\left[e^{(\epsilon-\mu)/T}-1\right]^2}\left(\frac{\epsilon-\mu}{T^2}+\frac{\mu^\prime}{T}\right)d\epsilon,$$
\end{widetext}
is discontinuous at $T=T_0$, i.e.,
\begin{eqnarray}
[C]&=&A\frac{[\mu^\prime]}{T_0}\int_0^\infty\frac{\epsilon^{\alpha}e^{\epsilon/T}d\epsilon}
{\left(e^{\epsilon/T}-1\right)^2}=-N
\frac{\alpha^2\zeta\left(\alpha\right)}{\zeta\left(\alpha-1\right)}.
\label{Cdisc}
\end{eqnarray}

The case when  the inequality
\begin{equation}
1<\alpha\leq 2\label{cond3}\end{equation} is satisfied, demands
the separate treatment, because  $e^{\mu n/T}$ under the sum in
Eq.~(\ref{muviaN1}) cannot be expanded in powers of $\mu/T$ due to
the divergence of the summation over  $n$.  In due turn,  both the
cases of the strict inequality  $1<\alpha<2$ and the equality
$\alpha=2$ in (\ref{cond3}) should also be treated separately.
Depending on the number of space dimensions  $D=1$, 2, and 3, the
strict inequality is satisfied for the power $\gamma$ in the
intervals $2/3<\gamma<2$, $\gamma>2$, and   $|\gamma|>6$,
respectively.

First, let us consider the case of the strict inequality
$1<\alpha<2$. At small $\mu$, the dominant contribution to the sum
in Eq.~(\ref{muviaN1}) comes from large $n$, so the summation can
be approximated by the integration with the help of the formula
due to Euler and Maclaurin \cite{LL5}:
\begin{eqnarray}
\sum_{n=1}^\infty f(n+a)&\approx&\int_a^\infty
f(x)dx-\frac{1}{2}f(a)-\frac{1}{12}
f^\prime(a).\label{EulerMac}\end{eqnarray}To this end one should
write
$$\sum_{n=1}^\infty f(n)=\sum_{l=0}^\infty f(l+1)=f(1)+\sum_{l=1}^\infty f(l+1),$$ and apply
(\ref{EulerMac}) in the particular case  $a=0$. Upon neglecting
the non-singular terms of the order of $\mu/T$, one finds that the
dominant contribution to the equation relating the chemical
potential $\mu$ to the particle number $N$ is thus represented in
the form
\begin{widetext}
\begin{eqnarray}
N&=&N\left(\frac{T}{T_0}\right)^{\alpha}\left[1+\frac{1}
{\zeta\left(\alpha\right)}\int_1^\infty\left(e^{x\mu/T}-1\right)
\frac{dx}{ x^{\alpha}}\right]\approx
N\left(\frac{T}{T_0}\right)^{\alpha}\left[1-\frac{1}
{\zeta\left(\alpha\right)\left(\alpha-1\right)}
\left(\frac{|\mu|}{T}\right)^{\alpha-1}
\Gamma\left(2-\alpha,\frac{|\mu|}{T} \right)\right],
\label{Nlong}\end{eqnarray}
\end{widetext}
where
\begin{eqnarray}
\Gamma(\alpha,x)&=&\int_x^\infty
e^{-z}z^{\alpha-1}dz=\Gamma(\alpha)-\nonumber\\&&
\sum_{n=0}^\infty\frac{(-1)^nx^{\alpha+n}}{n!(\alpha+n)},\label{Gammaalx}\end{eqnarray}
is the incomplete gamma function \cite{ryzh}, and $x=|\mu|/T\ll1$.
Because $x$ is small, one can keep  only the term with $n=0$ in
the sum over $n$  in Eq.~(\ref{Gammaalx}). As a result, after
keeping the terms of the lower order in the ratio $\mu/T$, one
obtains the first-order  expression for the nonzero value of
$\mu(T)$ for $T>T_0$:
\begin{equation}
\mu(T)\approx-T_0\left[\frac{\alpha\left(\alpha-1\right)
\zeta\left(\alpha\right)(T-T_0)}
{T_0\Gamma\left(2-\alpha\right)}\right]^{\frac{1}{\alpha-1}}.\end{equation}
This expression shows that the chemical potential is continuous at
$T=T_0$. Indeed, $\mu(T)$ vanishes at $T\leq T_0$ with all its
derivatives. Since, in the present case, $1/(\alpha-1)>0$, then,
at $T>T_0$, the expression $(T-T_0)^{\frac{1}{\alpha-1}}$ vanishes
at $T\to T_0$, too. As for the derivatives of the chemical
potential over temperature, the evaluation of the order $k$
derivative can be expressed as follows
\begin{eqnarray*}
\mu^{(k)}(T)&\equiv&\frac{\partial^k\mu}{\partial
T^k}=-\frac{1}{T_0^{k-1}}\left[\frac{\alpha(\alpha-1)}
{\Gamma(2-\alpha)}\right]^{1/(\alpha-1)}\times\nonumber\\&&
\left(\frac{T-T_0}{T_0}\right)^{1/(\alpha-1)-k}
\prod_{l=0}^{k-1}\left(\frac{1}{\alpha-1}-l\right),
\end{eqnarray*}
which shows that, provided the condition
\begin{equation}
\alpha=1+\frac{1}{k},\label{alk}\end{equation}is fulfilled, where
$k\geq2$, the $k$-th order  derivative of the chemical potential
as a function of temperature is discontinuous at   $T=T_0$. The
magnitude of the  discontinuity is
\begin{equation}
[\mu^{(k)}]=-\frac{k!}{T_0^{k-1}}\left[\frac{(k+1)\zeta(1+1/k)}
{k^2\Gamma(1-1/k)}\right]^k.\label{mukdisc}\end{equation} Since,
in the present case, $\alpha<2$,  the case $k=1$ is beyond the
treatment. As will be clear later on, when $\alpha=2$, the
temperature behavior of the chemical potential becomes
non-analytical at $T=T_0$.

Let us turn to the case $\alpha=2$.  Note that this equality takes
place, if, at the space dimension $D=1$, 2, 3, the power of the
coordinate dependence of the trapping potential $U(r)\propto
r^\gamma$ equals, respectively, $\gamma=2/3$, 2, 6. It is the
situation $D=2$, $\gamma=2$ which was realized recently in the
experiments with the Bose -- Einstein condensation of photons in
the cavity \cite{Kla}. The chemical potential at $T>T_0$ can be
found from Eq.~(\ref{muexact}), with  $\alpha=2$:
\begin{equation}
\left(\frac{T_0}{T}\right)^2=\frac{{\rm
Li}_2\left(e^{\mu/T}\right)}{{\rm Li}_2(1)}\approx\frac{{\rm
Li}_2\left(1-|\mu|/T\right)}{{\rm
Li}_2(1)}\label{muexact2},\end{equation} The corresponding special
function is now Li$_2(x)$. Its definition and some properties
necessary for the present treatment, are listed in the Appendix.
Using Eq.~(\ref{Li2eps}) with $\epsilon=|\mu|/T$,  the equation
for the determination of $|\mu|$ at $0<T-T_0\ll T_0$, reads
$$\frac{|\mu|}{T}\ln\frac{|\mu|}{T}\approx-\frac{\pi^2(T-T_0)}{3T_0},$$
whose solution, with the logarithmic accuracy, can be represented
in the form
\begin{equation}
\mu(T)\approx-\frac{\pi^2(T-T_0)}{3\ln\frac{3T_0}{\pi^2(T-T_0)}}.\label{muapprox}
\end{equation}
One can see that, in the experimentally accessible case   $D=2$,
$\gamma=2$ \cite{Kla},  the temperature dependence $\mu(T)$ is
non-analytical at $T=T_0$. The differentiation over temperature
shows that the chemical  potential and its first derivative are
continuous but the second derivative has an infinite discontinuity
at the critical temperature. Hence the temperature derivative of
the heat capacity is discontinuous.

Let us comment on the the thermodynamics of the Bose gas in the
limiting cases $\gamma=0$ or $\gamma\to\infty$. The first one is
reduced to the constant external potential $U(r)=U_0=$const. In
this case, the thermodynamical properties are meaningful, if one
restricts the system to the finite $D$-dimensional volume. Since
the effect of the potential $U(r)=$const can be removed by the
change of the energy reference point, the thermodynamical
properties at $\gamma=0$ are the same as in the case of free
bosons. Curiously, but the seemingly opposite case
$\gamma\to\infty$ corresponds exactly to the same case of the free
Bose gas. The only difference is that the effect of the finite
volume is reached naturally. See Eq.~(\ref{gamminf}). The behavior
of the BEC fraction $N_0(T)$ and the  chemical potential $\mu(T)$
near the BEC transition point $T_0$ can be inferred from the
results obtained in the preceding section by making the
replacement $\alpha\to D/2$. In particular,
$$N_0(T)=N\left[1-\left(\frac{T}{T_0}\right)^{D/2}\right],$$
which coincides with the results of \cite{achar}. For the ideal
Bose gas, BEC is possible when $D>2$. If  $D>4$, the chemical
potential is continuous at $T=T_0$, but its derivative over $T$
has the discontinuity of the magnitude
$$[\mu^\prime]=-\frac{D\zeta(D/2)}{2\zeta(D/2-1)}.$$
If $D/2=1+1/k$, where $k=2,3,\cdots$, that is, $D=3,8/3,\cdots$,
then the second, third, $\cdots$ derivative of the chemical
potential is discontinuous. In fact, one finds from
(\ref{mukdisc}), that, in the three-dimensional space, the
magnitude of the discontinuity is
$$[\mu^{\prime\prime}]=-\frac{9\zeta^2(3/2)}{8\pi
T_0}\approx-\frac{2.44}{T_0}.$$This coincides with the textbook
result. Finally, if $D=4$, then $\mu(T)$ for $T>T_0$ is given by
the expression (\ref{muapprox}), the same as in the case of the
oscillatory external potential in the two-dimensional space.

\section{Discussion and conclusion}
\label{disc}~

The temperature behavior of the chemical potential as the function
of the temperature  is important from the point of view of
establishing the kind of the phase transition, if such a
transition indeed takes place. There is a well-known Ehrenfest
classification of the phase transitions according to which the
transition is of the first order, if the energy (in particular,
the chemical potential) is discontinuous, of the second order, if
the derivative of the energy over temperature (the heat capacity)
is discontinuous, etc. If one treats the Bose -- Einstein
condensation as the phase transition, then the results of the
preceding section help to establish the kind of this transition.
One can see, that in all cases except the one with $\alpha=2$, the
BEC transition is indeed the phase transition of some particular
kind. Specifically, when $\alpha>2$, the phase transition is of
the second kind [see Eq.~(\ref{Cdisc})], when $\alpha=1+1/k$,
where $k\geq2$, the transition is of the $(k+1)$-th kind [see
Eq.~(\ref{mukdisc})]. But in the singular case of $\alpha=2$,
which is realized in $D=4$ for the bosons in zero external
potential (unphysical case), and in the two-dimensional oscillator
(physically realized in \cite{Kla}), the BEC transition is the
phase transition of the third kind.

The kind of the phase transition could be established in the
caloric experiments, i.e. via the measurement of the temperature
dependencies of various thermodynamical quantities. Is the
crossover character (\ref{muapprox}) of the BEC transition
realized or not,  was not studied in the experiment \cite{Kla}
with the $D=2$ Bose gas of photons in the oscillator potential
$U(r)\propto r^2$. The fact is that the Bose -- Einstein
condensation in the experiment \cite{Kla} took place at the fixed
room temperature. BEC was manifested as the macroscopic occupation
of the state with the energy $E_0=\hbar cs/D_0$, see
Eq.~(\ref{Edisp}), upon reaching some critical pump power of the
laser corresponding to $N\sim10^4$ photons in the cavity. This
type of the experiment  is not the caloric one.

To summarize, the power $\gamma$ of the trapping spherically
symmetric potential, $U\propto r^\gamma$,  for dimension of the
space $D$ which admits the Bose -- Einstein condensation of the
ideal Bose gas, is found.  The degree of singularity of the
chemical potential and other thermodynamic functions appears to be
very different for different space dimensions $D$ and different
power $\gamma$ of the coordinate dependence of the trapping
potential encoded in the quantity $\alpha=D/2+D/\gamma$. In
particular, the case $\alpha=2$ corresponding to $D=2$ and
$\gamma=2$,  accessible through  the recent experiment \cite{Kla}
with photons in the cavity, results in the non-analytic behavior
of the temperature dependence of the chemical potential and other
thermodynamic functions. The results obtained in the paper, could
be possibly applied in the case of the hypothetical axionic BEC as
the dark matter \cite{Sik}, in some power-like spherical-symmetric
potentials in 3D,  realized in the self-gravitating distributions
of the above hypothetical  particles.

The author thanks an anonymous referee for the careful refereeing
and the constructive remarks and  suggestions. I am grateful to
Prof. Martin Weitz for correspondence which helped to reveal  the
error in the first draft of the manuscript.

After this manuscript was accepted for publication I have been
informed by Prof. V.~I.~Yukalov  about his results \cite{yukalov}
on similar subject.

\appendix
\section{Polylogarithm function} \label{app} ~

The polylogarithm function ${\rm Li}_\alpha(z)$ is defined by the
following integral and series representations \cite{prud}:
\begin{equation}
{\rm
Li}_\alpha(z)=\frac{z}{\Gamma(\alpha)}\int_0^\infty\frac{x^{\alpha-1}dx}{e^x-z}
=\sum_{k=1}^\infty\frac{z^k}{k^\alpha}.\label{Li}
\end{equation}
This function is defined at $|z|\leq1$, and $\alpha>1$. As is
clear from this definition, ${\rm Li}_\alpha(1)=\zeta(\alpha)$. In
particular, another equivalent representations for ${\rm Li}_2(z)$
is used in the paper:
\begin{equation}
{\rm
Li}_2(z)=-\int_0^z\ln(1-t)\frac{dt}{t}.\label{Li2}\end{equation}
To prove this representation, one should use the series expansion
of $\ln(1-t)$ \cite{gradlog}. Then
$$\frac{\ln(1-t)}{t}=-\sum_{k=1}^\infty\frac{t^{k-1}}{k},$$ and
$$-\int_0^z\ln(1-t)\frac{dt}{t}=\sum_{k=1}^\infty\frac{z^k}{k^2}={\rm
Li}_2(z).$$The expansion of ${\rm Li}_2(z)$ near $z=1$ used in the
body of the paper, is
\begin{equation}
{\rm
Li}_2(1-\epsilon)\approx\frac{\pi^2}{6}+\epsilon\ln\epsilon,\label{Li2eps}\end{equation}
where $\epsilon\ll1$. Indeed, one has the following chain of
equations:
\begin{eqnarray*}
{\rm
Li}_2(1-\epsilon)&=&-\int_0^{1-\epsilon}\ln(1-t)\frac{dt}{t}=\nonumber\\&&-\left(\int_0^1-\int_{1-\epsilon}^1\right)
\ln(1-t)\frac{dt}{t}=\nonumber\\&&{\rm
Li}_2(1)+\int_{1-\epsilon}^1\ln(1-t)\frac{dt}{t}=\nonumber\\&&{\rm
Li}_2(1)+\int_0^\epsilon\frac{du\ln
u}{1-u}\approx\nonumber\\&&\frac{\pi^2}{6}+\epsilon\ln\epsilon.
\end{eqnarray*}
The approximate equality has the logarithmic accuracy, and  the
relation $\zeta(2)=\pi^2/6$ is taken into account.



\begin{thebibliography}{99}
\bibitem{Pit}
F.~Dalfovo, S.~Giorgini, L.~P.~Pitaevskii, and S.~Stringari, Rev.
Mod. Phys. {\bf71}, 463 (1999).
\bibitem{Bag}
V.~Bagnato, D.~E.~Pritchard, and D.~Kleppner, Phys. Rev. A{\bf35},
4354 (1987).
\bibitem{Mull}
W.~J.~Mullin.  J. Low Temp. Phys. {\bf106}, 615 (1997).
\bibitem{achar}
A.~Acharyya and  M.~Acharyya, Acta Physica Polonica B{\bf 43},
1805 (2012); arXiv:1208.4888v1 [cond-mat.stat-mech].
\bibitem{Kla}
J.~Klaers, J.~Schmitt, F.~Vewinger, and M.~Weitz, Nature, {\bf
468}, 545 (2010).
\bibitem{Sik}
P.~Sikivie and Q.~Yang, Phys. Rev. Lett. {\bf103}, 111301 (2009).
\bibitem{ryzh}
I.~M.~Gradshtein and I.~S.~Ryzhik, \textit{Tables of integrals,
sums, series, and products}, (FizMatLit, Moscow, 1963).
\bibitem{grad1}
See \cite{ryzh}, the formulae 8.380(1) and 8.384(1).
\bibitem{LL5}
L.~D.~Landau and E.~M.~Lifshits, \textit{Statistical Physics, part
I}, (Nauka Publishers, Moscow, 1965, in Russian), Chapter V.
\bibitem{prud}
A.~P.~Prudnikov, Yu.~A.~Brychkov, and O.~I.~Marichev,
\textit{Integrals and seris. Elementary functions}, (Nauka
Publishers, Moscow, 1981), p.790.
\bibitem{gradlog}
See \cite{ryzh}, the formula 1.511.
\bibitem{yukalov}
V.~I.~Yukalov, Phys. Rev. A{\bf72}, 033608 (2005).
\end{thebibliography}
\end{document}